\documentclass[11pt]{article} 
\usepackage[normalem]{ulem}
\usepackage{graphicx}
\usepackage{hyperref}
\usepackage{amssymb}
\usepackage{authblk}
\usepackage{lineno}
\usepackage{color}

\date{\today}
\title{Monte Carlo simulations for the ANTARES underwater neutrino telescope }
\author[1,2]{A.~Albert}  
\author[3]{M.~Andr\'e}  
\author[4]{M.~Anghinolfi}  
\author[5]{G.~Anton}  
\author[6]{M.~Ardid}  
\author[7]{J.-J.~Aubert}  
\author[8]{J.~Aublin}  
\author[8]{B.~Baret}  
\author[9]{S.~Basa}  
\author[10]{B.~Belhorma}  
% ...
\author[  ]{                                                                            (The ANTARES Collaboration)}

\affil[1]{\scriptsize{Universit\'e de Strasbourg, CNRS,  IPHC UMR 7178, F-67000 Strasbourg, France}  }  
\affil[2]{\scriptsize{Universit\'e de Haute Alsace, F-68200 Mulhouse, France}  }  
\affil[3]{\scriptsize{Technical University of Catalonia, Laboratory of Applied Bioacoustics, Rambla Exposici\'o, 08800 Vilanova i la Geltr\'u, Barcelona, Spain}  }  
\affil[4]{\scriptsize{INFN - Sezione di Genova, Via Dodecaneso 33, 16146 Genova, Italy}  }  
\affil[5]{\scriptsize{Friedrich-Alexander-Universit\"at Erlangen-N\"urnberg, Erlangen Centre for Astroparticle Physics, Erwin-Rommel-Str. 1, 91058 Erlangen, Germany}  }  
\affil[6]{\scriptsize{Institut d'Investigaci\'o per a la Gesti\'o Integrada de les Zones Costaneres (IGIC) - Universitat Polit\`ecnica de Val\`encia. C/  Paranimf 1, 46730 Gandia, Spain}  }  
\affil[7]{\scriptsize{Aix Marseille Univ, CNRS/IN2P3, CPPM, Marseille, France}  }  
\affil[8]{\scriptsize{Universit\'e de Paris, CNRS, Astroparticule et Cosmologie, F-75013 Paris, France}  }  
\affil[9]{\scriptsize{Aix Marseille Univ, CNRS, CNES, LAM, Marseille, France }  }  
\affil[10]{\scriptsize{National Center for Energy Sciences and Nuclear Techniques, B.P.1382, R. P.10001 Rabat, Morocco}  }  
% ...
\begin{document}
\maketitle 

\thanks{\href{mailto:annarita.margiotta@unibo.it}{annarita.margiotta@unibo.it}}

\begin{abstract}
Monte Carlo simulations are a unique tool  to check the response of a detector and to monitor its performance.
For a deep-sea neutrino telescope, the  variability of the environmental conditions  that can affect the behaviour of the data acquisition system must be considered,  in addition to a reliable description of the active parts of the detector and of the  features of physics events, in order to produce a realistic set of simulated events. In this paper, the software tools used to produce neutrino and cosmic ray signatures in the telescope and the strategy developed to represent the time evolution of the natural environment and of the detector efficiency are described.
\end{abstract}
\section{Introduction} \label{sec:intro}
Interest in high-energy (HE) neutrino astrophysics ({E$_\nu>$100\,GeV}) has been rapidly increasing in the last ten years. The  measurement of a diffuse cosmic flux of neutrinos reported by the IceCube Collaboration \cite{bib:ic, bib:ic_comb} and the possible identification of a neutrino source \cite{bib:ic_source1} have further enhanced the importance of neutrino astronomy.

HE neutrinos can be created in the interaction of cosmic ray (CR) protons or nuclei in the proximity of their astrophysical sources and  travel undeflected and unabsorbed being neutral and weakly interacting particles. They thus represent the ideal probe to explore the far and energetic Universe. The main drawback of this feature is their small interaction cross section, which makes neutrino detection challenging.

Neutrino telescopes  can be built by instrumenting a large volume of water or ice with a three-dimensional array of photosensors. 
Neutrino interactions taking place in the vicinity of the apparatus can be observed by detecting Cherenkov photons emitted along the path of the relativistic charged particles that are produced. 

A cosmic neutrino signal can be identified using different approaches: looking for a directional excess coming from resolved sources or for an excess of very high-energy events (diffuse flux) either emitted by an ensemble of unresolved sources or due  to the propagation of CRs  through the Universe; in a multi-messenger context, searching for space/time coincidences of neutrino observations with electromagnetic probes over the whole spectrum, CRs or gravitational waves. Whichever is the selected approach, the discrimination between the physics signal and the expected background is a crucial point  to perform an accurate statistical analysis. Monte Carlo (MC) simulations play an essential role in the comprehension of the detector response to the different sources of optical signals: incident neutrinos, atmospheric muons, and natural background radiation. 

ANTARES is an underwater neutrino telescope \cite{bib:antares} whose main goal is the exploration of the Southern sky, searching for HE neutrino sources,  particularly in the region of the Galactic plane and centre, for which the detector has a privileged field of view \cite{bib:extps, bib:galridge}. Thanks to its location in the Mediterranean Sea and to the good angular resolution, constraints have been put on  various hypotheses  about the origin of the cosmic signal reported by the IceCube Collaboration \cite{bib:allflavour,bib:antic,bib:icblaz}.  Moreover, various results have been published concerning the indirect search for dark matter  \cite{bib:dmsecl, bib:dmsun, bib:dmgal, bib:dmearth, bib:dmlast}.  
An intense program of multi-messenger research is also underway, looking for neutrino events together with other astrophysical signals. Specifically, the ANTARES Collaboration has analysed data  for possible presence or coincidences with  gravitational waves \cite{bib:multimgw1, bib:multimgw2, bib:multimgw3, bib:multimgw4, bib:multimgw5, bib:multimgw6, bib:multimgw7}, gamma-ray bursts \cite{bib:multimgrb1, bib:multimgrb2}, ultra-high energy cosmic rays \cite{bib:multimcr} and emissions in different ranges of the electromagnetic spectrum \cite{bib:multimradioburst1, bib:multimradioburst2, bib:multimradioburst3, bib:multimradioburst4, bib:multimradioburst5}.

In this paper, the main aspects of the simulation software chain developed within the  Collaboration and used in almost all publications  are discussed. In Sec. \ref{sec:antares}  the ANTARES detector and its main physics goals are presented. 
Then, the full simulation procedure is reviewed and its components are described: the software used to generate physics events in Sec. \ref{sec:ev_gen}, the simulation of emission and  propagation of the Cherenkov photons  in Sec.  \ref{sec:km3}, and, in Sec. \ref{sec:daq_sim},  the simulation of the detector response and the reproduction of the data stream.  In Sec. \ref{sec:rbr} the strategy used to follow the time evolution of the data acquisition conditions is described. Finally, some general conclusions are drawn in Sec. \ref{sec:conclusion}.
\section{The ANTARES neutrino telescope} \label{sec:antares}
The ANTARES neutrino detector \cite{bib:antares} is installed at a depth of about 2.5 km under the Mediterranean Sea. It is located at (42$^\circ$\,48$^\prime$\, N, 6$^\circ$\,10$^\prime$\,E), 40\,km from Toulon, France. The detector consists of twelve lines, each 450\,m long,  anchored to the seabed and kept taut by submerged top buoys. Lines are horizontally-spaced, on average, by  about 70\,m. Each line but one carries 25 storeys spaced by 14.5\,m, which are assembled structures supporting each one  3  optical modules (OMs) \cite{bib:antares_om}, pressure-resistant glass spheres housing a 10-inch photomultiplier (PMT) \cite{bib:antares_pmt}, and a titanium container holding electronic boards and positioning devices. On the remaining line, the top five storeys are substituted with acoustic receivers \cite{bib:amadeus}. ANTARES has been taking data continuously since 2008. 

The optical modules  collect light emitted along the path of   charged particles produced in neutrino interactions  and atmospheric muons. In addition, environmental sources -- namely luminous bacteria, macro-organisms and the decay products of $^{40}$K -- contribute to the overall amount of detected signals \cite{bib:current1, bib:current2} and are the main cause of the environmental optical background.

When  photons of any origin impinge on the optical sensor photocathode, a signal can be measured at the  anode  and converted into digital format by the front-end electronics boards \cite{bib:antares_electronics}, recording time, position and charge, and storing information  in what is called a \textit{hit}.
All hits  with a charge above a minimum threshold of 0.3 photoelectrons (p.e.) are transmitted to the shore and processed  with dedicated trigger algorithms to identify potentially interesting events that are  stored on disk \cite{bib:antares_daq}. Different triggers are applied, based on local coincidences, where a local coincidence is   the occurrence of  hits on two separate OMs in the same storey within a 20~ns window or a large amplitude single hit, typically larger than 3 p.e. Two main standard triggers are defined. The first is a directional scan logic trigger, which requires five causally connected coincidences  within a triggering time window of 2.2$~\mu$s, which is roughly the time required for a muon to travel through the detector. The second is a cluster logic trigger, which requires a combination of two local coincidences in adjacent or next-to-adjacent storeys within 100 ns or 200~ns respectively. When a trigger occurs, all hits found in a predefined time window, usually 2.2$~\mu$s before and after the first and the last hit of the event, are recorded.  

The  result of the MC simulation chain consists of several sets of events whose format is identical to real data and  ready to be reconstructed  and analysed with the  programs used for the data stream collected with the detector.

The simulation procedure is subdivided into different steps:
\begin{enumerate}
\item Generation of the physics events: the kinematic information of each detectable particle is generated in the proximity of the detector. 
\item Tracking: particles are propagated through the detector and the Cherenkov photons are simulated and propagated to the optical modules.
\item Data acquisition: a simulated data stream is produced using the simulation of the PMT response, the signal digitisation and adding the optical background. Finally, filtering algorithms are applied, identical to those used for  the real data stream. The  evolution of the detector efficiency as a function of time is also accounted for at this step.
\end{enumerate}
In the following, the focus is set on the simulation of neutrino interactions of any origin (atmospheric, cosmic,  from dark matter annihilation,...) and of atmospheric muons, which represent the bulk of the events detected in a neutrino telescope. Specific algorithms have also been developed to simulate the passage of exotic particles \cite{bib:RS}, for example  magnetic monopoles \cite{bib:mm1, bib:mm2}.

\section{Event generation} \label{sec:ev_gen}
\begin{figure}
  \centering
  \includegraphics[width=0.8\textwidth]{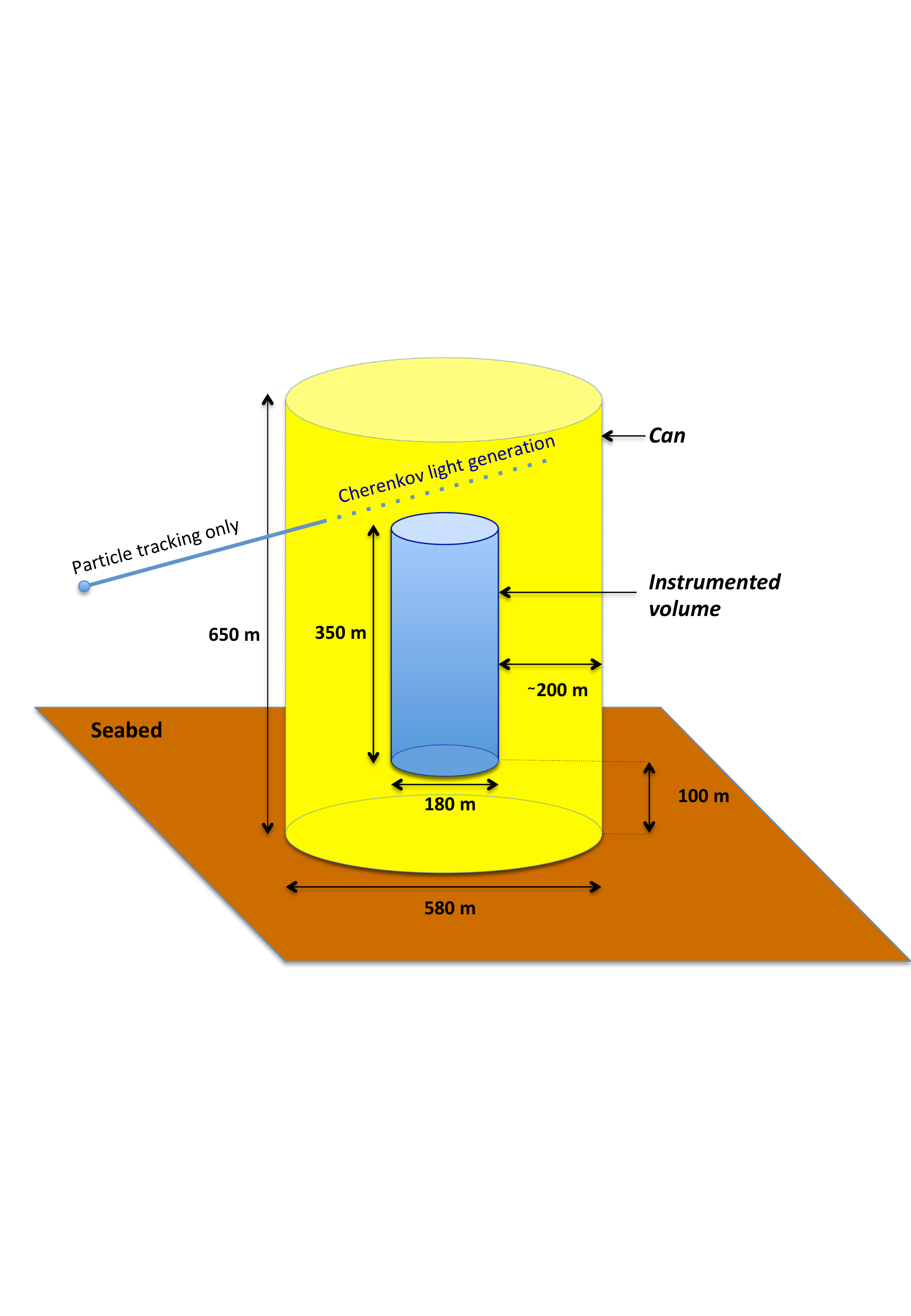}
  \caption{Schematic view of the ANTARES can (in yellow), anchored to the seabed (in brown) and containing the detector instrumented volume (in blue).}
  \label{fig:can}
\end{figure}

In this first step, the energy, direction and starting position  of each charged particle  are defined. All charged particles that can induce Cherenkov photons with high probability to arrive at the sensitive components of the detector are considered. The sensitive volume of the detector is called the \textit{can}. The can is  a cylinder bounding the water region that hosts the PMTs (instrumented volume) extended by a quantity depending on the optical water properties, namely the attenuation length of  light in water \cite{bib:riccobene}. It defines the  volume where the Cherenkov light is generated in the  simulation (Figure \ref{fig:can}). Cherenkov photons produced out of this region have a low probability to reach a PMT. Therefore, outside this volume, only energy losses of long tracking particles (i.e. muons and taus) are considered. 
\subsection{Neutrinos} \label{sec:nu_gen}

Charged current (CC) and neutral current (NC) interactions of neutrinos of any flavour, from the sub-GeV energy range up to 10$^9$ GeV, are generated with the GENHEN (GENerator of High-Energy Neutrinos)   
code.  Hereafter,  though the simulation algorithms take into account the peculiarities of the neutrino and anti-neutrino interactions, ``neutrino" will refer to  both.
 All relevant processes are considered, including the so-called ``Glashow resonance" \cite{bib:glashow} in the simulation of CC interactions of electron anti-neutrinos. Deep inelastic scattering, dominant at high energy, is simulated using the LEPTO package  \cite{bib:lepto}. Above 10 TeV,  an extrapolation of the model is used to calculate cross sections and  interaction kinematics up to 10$^9$ GeV. The  CTEQ6D \cite{bib:cteq} parton distribution functions are used.

Individual neutrinos are injected into the code, usually according to an energy spectrum in the form of  a power law $dN/dE \propto E^{-\gamma}$, where $\gamma$ is chosen by the user in order to have an adequate statistical significance  across the considered energy range.  Afterwards, events can be weighted according to a specific flux model, depending on the analysis to be  performed: studies of atmospheric neutrinos, point-like sources, diffuse fluxes, etc.

A generation volume (V$_{\rm gen}$) is considered: within this volume, whose size depends on the neutrino interaction type and on the neutrino flavour and energy, the position of the neutrino interaction vertex and its direction are randomly drawn, following a  uniform distribution (the dimensions of this volume are always much less than the neutrino interaction length).
Every neutrino is considered as interacting within this volume, and   secondary interactions are produced at the interaction vertex. In order to optimise  computer processing time, each neutrino flavour and  interaction type  (NC or CC)  is  treated separately. 

For NC  or $\nu_{e}$ CC events, initial interaction products are  simulated  at the vertex, with short-living particles decayed according to the  LEPTO prescriptions. The electron produced in $\nu_e$ CC interactions induces an electromagnetic shower; the charged secondary hadrons induce a hadronic shower. The longitudinal extension of either hadronic or electromagnetic showers is typically a few metres in water and the  V$_{\rm gen}$ is normally coincident with the can. 

If a $\nu_{\mu}$ or $\nu_{\tau}$ CC event is considered, a larger V$_{\rm gen}$ than the can is usually required because the long track of the leading charged lepton can be detected even when the vertex is far away from the can. In this case the  lepton is propagated if its direction intercepts the can, evaluating the energy loss occurring during the path. The distance between the vertex and the can and, consequently, the size of V$_{\rm gen}$ are calculated  according to the energy dependent  range of the leading charged lepton,  assuming it takes all neutrino energy. 
For  $\nu_{\tau} $ CC interactions different V$_{\rm gen}$ are defined corresponding to different  $\tau$  decay channels. When a muon is present  in the final state of the $\tau $ decay (Branching Ratio (BR)  $\simeq$ 17.4\%) \cite{bib:pdg} the V$_{\rm gen}$ is defined by the total  range of the two particles, i.e.  the sum of the $\tau$ and of the $\mu$ ranges. Otherwise, BR $\simeq$ 82.6\%,  the V$_{\rm gen}$ is almost coincident with the can if E$_\nu <$ 1 PeV; at higher energy, being the $\tau$ range larger than    50 m, a V$_{\rm gen}$ exceeding the can size, and depending on the actual $\tau$ range, must be considered.

Interactions can occur either in water close to the can volume or in the rock below the detector. The two different media are considered, assuming an isoscalar target with appropriate density.

Neutrino interactions are treated differently depending on whether the vertex is internal or external to the can.
When the interaction occurs within the can volume, the kinematical information of all final-state products (charged leading lepton, if present, and  charged particles in hadronic and electromagnetic showers) is stored and becomes the input to the program simulating the Cherenkov light. If the vertex is outside the can, only the long tracking particles, muons and taus, are considered for the following steps. The event is discarded if the distance between the vertex and the entry point at the can is larger than the maximum lepton range.

The energy interval chosen  by the user is subdivided into a user-defined number of equal divisions in log$_{10}(E_\nu)$.  
Usually,  the propagation through the Earth is not considered for  $\nu_\mu$ and $\nu_e$ and the neutrino energy at the interaction is the energy of the neutrino when it enters the Earth, sampled according to the user defined spectrum. The probability of Earth absorption is accounted for in the final weight.
For $\nu_\tau$ interactions the propagation through the Earth is fully considered and the $\nu_\tau$ regeneration effect is evaluated. The regeneration effect implies to consider the $\nu_\tau  \rightarrow \tau \rightarrow \nu_\tau $ decay chain, producing a $\nu_\tau$ with lower energy in the final state. 
\subsection{Atmospheric muons} \label{sec:mu_gen}
\begin{figure}
  \centering
  \includegraphics[width=1.0\textwidth]{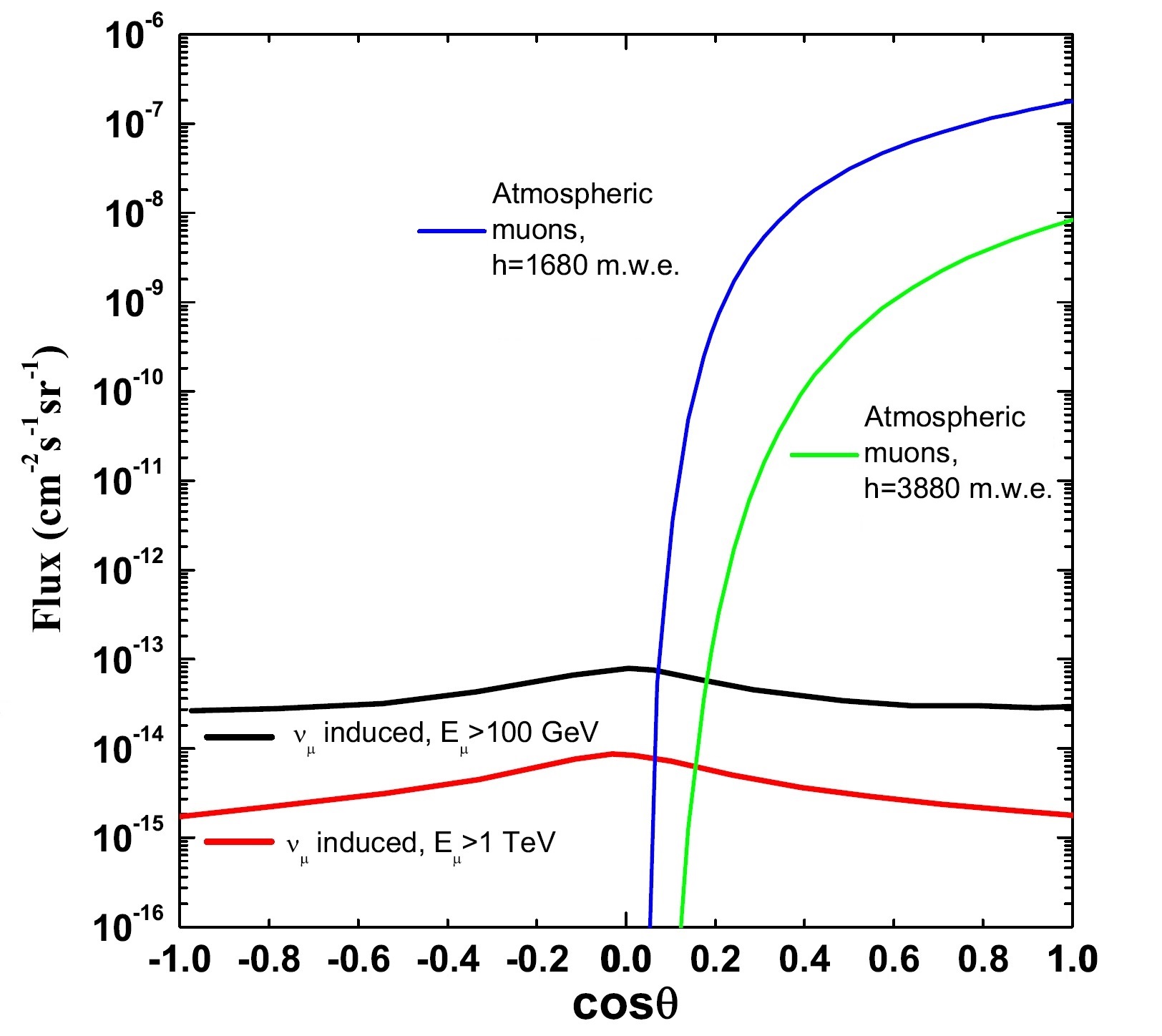}
  \caption{Atmospheric muon flux, evaluated following the parameterisation in Ref. \cite{bib:mupage_par} at two different depths (1680 and 3880 m water equivalent) compared to the flux of muons due to atmospheric muon neutrinos considering two different muon energy thresholds. The neutrino flux is calculated according to Ref. \cite{bib:nu_agrawal} and the plot taken from Ref. \cite{bib:spurio_chiarusi}.  }
  \label{fig:atmunu}
\end{figure}
Atmospheric muons produced in the interactions of CRs in the upper atmosphere represent the majority of the reconstructed events in the detector.  Despite the  shielding effect of the overlying water, a significant flux of high-energy atmospheric muons reaches the active volume of the detector (see Figure \ref{fig:atmunu}). Since the rate of such events with respect to  neutrino events is large, even selecting events reconstructed  as upward-going by the tracking algorithms and  with high quality criteria, a significant contamination due to atmospheric muons remains  \cite{bib:lastpoint}. An accurate simulation of atmospheric muons is required in order to estimate properly the background in the final sample of each data analysis.
In addition, atmospheric muons provide an almost constant and stable flux of particles and the  comparison between the behaviour of the detector and its expected performance allows to monitor the time evolution of the detector efficiency. They are also used for time calibration of the detector, as described in Ref.  \cite{bib:time_cal}.
   
The flux of atmospheric muon bundles  at the detector can be reproduced using a complete simulation of the atmospheric showers induced by the arrival of a primary cosmic ray or evaluating the underwater muon flux with a set of parametric formulae.
Both strategies have been considered: CORSIKA  \cite{bib:corsika} has been used for the full simulation and MUPAGE  \cite{bib:mupage} for the parameterised approach.

CORSIKA  is a widely used software which follows and tracks all particles produced in the interactions of  primary cosmic rays with the atmospheric nuclei, performing a complete extensive air shower simulation, from the top of the atmosphere to  sea level. 
At the cost of  high computational time, the program allows a broad simulation flexibility, offering a large choice for  input parameters:  atmosphere models according to different seasons and geographical locations, parameterisations of the hadronic interaction, chemical composition and energy spectrum of the primary cosmic ray flux and inclusion of charmed particle production. The muon flux measured with the ANTARES detector both in its partial configuration, with 5 lines \cite{bib:5line}, and after its completion has been compared to the expectations obtained with the CORSIKA 6.2 version and the QGSJET.01 \cite{bib:qgsjet} description of the high-energy hadronic interactions. 

The bulk of primary cosmic rays arriving at the top of the atmosphere has been represented with five groups of nuclei: protons, He, the C, N and O group, the Mg and Si group and Fe, produced according to a power law $E^{-2}$ over an energy range between 1 and 10$^{5}$ TeV/nucleon and  zenith angles from 0$^{\circ}$ to 85$^{\circ}$. A total number of showers larger than 10$^{10}$ has been simulated. All muons from showers reaching the sea level with energies larger than about 500 GeV  are transported to the detector using the program MUSIC \cite{bib:music}, a 3-dimensional muon propagator accounting for the main processes of muon energy loss. The properties of each muon hitting the can surface are registered for  future processing.  Each event has a weight accounting for the  spectrum used in the generation. This allows the application of a reweighting procedure at the analysis stage to account for chemical composition of the primary cosmic rays. Different hypotheses for the primary cosmic ray composition have been considered  \cite{bib:5line}.

A faster alternative for atmospheric muon simulation used in ANTARES is the MUPAGE software \cite{bib:mupage,bib:mupage_par}. This package is based on a set of parametric formulae extracted  from  a  full simulation of events, tuned according to the results of the MACRO experiment at the Gran Sasso Laboratory \cite{bib:macro} and extrapolated under the sea. The software provides the angular and energy distribution of muons at different depths as a function of the muon bundle multiplicity. The usage of parametric formulae allows the fast production of a large number of Monte Carlo events at the can surface.   This approach lacks of flexibility in the definition of the input parameters related to the primary composition and interaction models. Despite this limitation and considering the large uncertainties on the description of the hadronic interactions at very high energies and on the cosmic ray composition, the multi-year experience with the ANTARES detector has shown that this fast parametric simulation produces a reliable estimate of the atmospheric muon background and allows an efficient evaluation of the time evolution of its performances. Comparisons between atmospheric muon data and MUPAGE parameterisation are available in almost all quoted articles with ANTARES results.

\section{Particles and light propagation} \label{sec:km3}
The Cherenkov photons  induced by high-energy muons and secondary charged particles while traveling through water at relativistic speed  are simulated using a dedicated software package, KM3. The simulation of the light production and its propagation is sampled from ``photon tables" that store the numbers and the arrival times of the photons onto the photocathode and the probability of PMT hits, considering different distances, positions and orientations of the OMs with respect to a muon track or shower. The package is a suite of three different  programs designed to accomplish different  tasks, step by step, using the output of a program as input to the following.

\subsection{Cherenkov light generation} 
\begin{figure}
  \centering
  \includegraphics[width=1.0\textwidth]{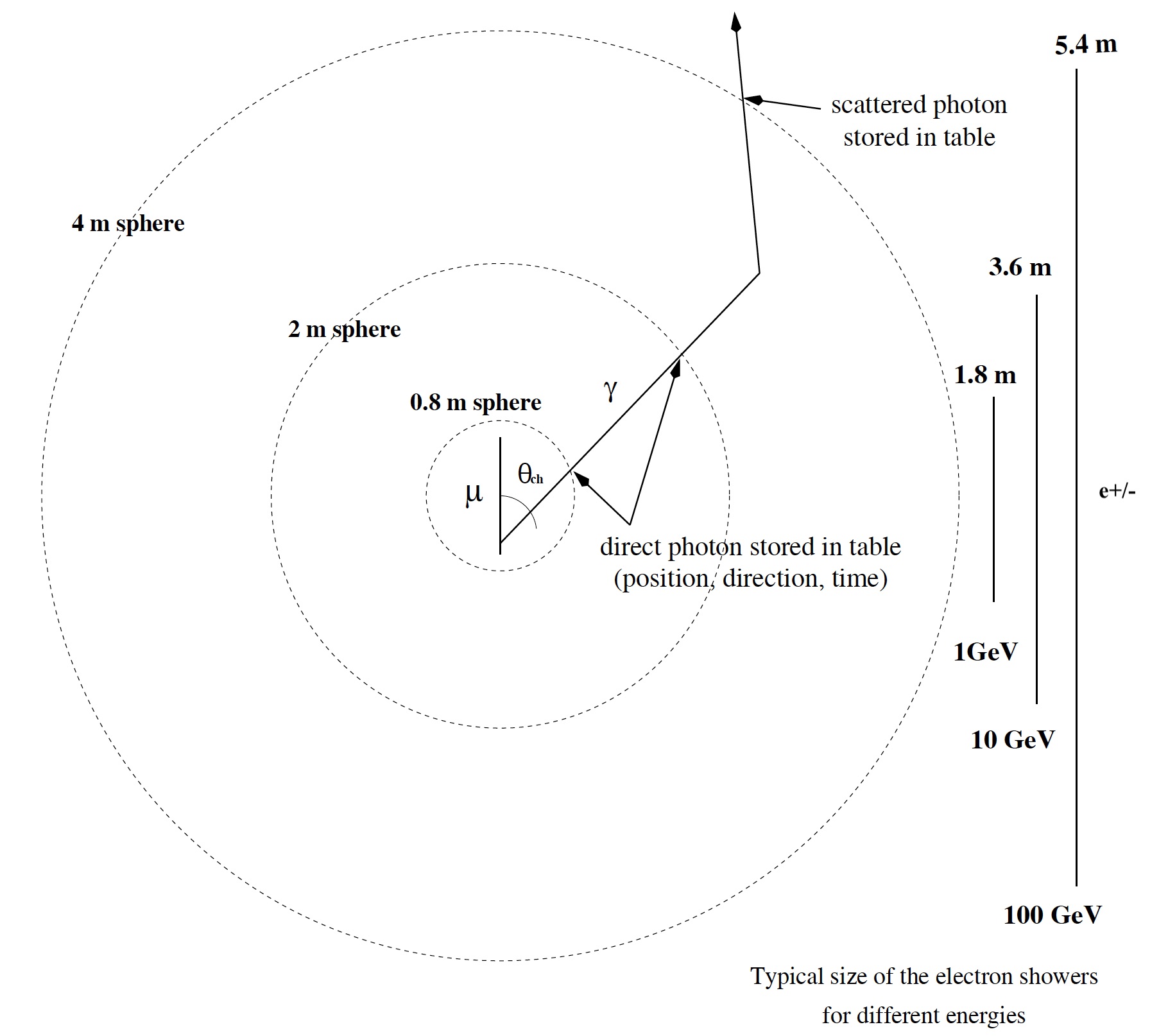}
  \caption{Graphical representation of the emission and propagation of Cherenkov photons induced by muons and electrons through the medium, crossing concentrical spheres at different distances.}
  \label{fig:km3_shell}
\end{figure}

The first program, called \textit{gen}, produces the ``photon fields". It performs  the Cherenkov light generation due to  muon or electron passage. Photons are tracked individually through water, until they leave the detector or are absorbed.
Wavelength dependent absorption and scattering are taken into account to evaluate their position, direction and arrival time at spherical concentrical shells of various, increasing radii around the light source. The original particle can be a muon or an electron. In the case of muons the photons produced by 1 m long track are considered. For electrons, the considered track length depends on  momentum. 
The number of concentrical shells can be modified by the user, extending the propagation distance of photons. The maximum radius used for ANTARES  simulations is set to 400 m. Figure \ref{fig:km3_shell} shows how the Cherenkov photons are produced and propagated through the medium in the case of a muon track segment. The symbol $\theta_{Ch} $ indicates the Cherenkov emission angle in water, about 42$^\circ$, between the track  and the photon directions. Information on both direct and scattered photons  at the boundaries of the shells, increasingly further away from the track, 
 is stored in a set of tables. Also the typical extension of an electromagnetic shower is shown for different energies of the electron.
Water absorption produces  a decrease of the total light impinging on an optical module,  reducing the  number of detected photons. Each scattering event changes the direction of the photons, increasing their path length with respect to direct, unscattered photons. Light scattering smears the arrival time distribution of photons and degrades  the angular resolution of the reconstructed parent neutrino direction. Both effects must be considered in the simulation to reproduce real conditions.
As input to the ANTARES simulation, the absorption length spectrum is taken from the measurements performed in several Mediterranean sites and reported in Ref. \cite{bib:riccobene}. These parameters are compatible with previous partial measurements (only ultraviolet and blue wavelengths were considered) performed at the ANTARES site 
 \cite{bib:antares_water}. Light scattering has been parameterised according to the model in Ref. \cite {bib:mobley}, using a  combination of Rayleigh and Mie scattering, with the ratio between  Rayleigh  and total scattering set to $\eta =0.17$.
\subsection{Hit production } The second program of the KM3 package, \textit{hit}, uses the photon fields produced with the  \textit{gen} code to evaluate the  probability  of registering a hit on the PMTs. 
As  input to the code the effective area of the OM is used. Its evaluation is based on a full GEANT \cite{bib:geant4} simulation and it is defined as the ratio of detected to incident photons multiplied by the cross-sectional area illuminated in a simulation. It corresponds approximately to the projected geometrical area of the photocathode multiplied by the quantum efficiency.  The nominal efficiency of the OMs is globally normalised according to the measurements performed in the laboratory dark room  \cite{bib:antares_om}.
The output of \textit{hit} is a set of   tables, containing the probability for photons induced by electrons of different energy and by relativistic muon track segments to produce a detectable signal on an OM.
\subsection{Particle propagation }  The  third part of the KM3 package, the program \textit{km3mc},  is dedicated to the propagation of the particles and of the light through the can volume.
KM3  is able to treat only the light induced by muons and  by  electromagnetic showers that can be either subshowers emitted along the muon path or direct high-energy electron cascades.  
Muons are transported  using the MUSIC package \cite{bib:music} considering  1 m long track segments, until the muon stops or leaves the detector. At each step all energy-loss processes  are considered and the muon energy-loss is treated as continuous, if the energy loss along the step is lower than 300 MeV/m, or discrete/stochastic, if the energy loss exceeds 300 MeV/m.  In the  case of continuous energy loss, the expected number of photons is extracted from the muon tables calculated with  the previous \textit{hit} program. For discrete energy losses,  an independent electromagnetic shower is assumed to be present  at a random position along the segment length and the number of photons is extracted from the corresponding electron tables. An energy dependent scaling of the amount of light is considered. Direct and scattered photons are both counted.
In the case of hadronic showers, a large number of charged particles is produced at the interaction vertex. Due to the high stochastic variability in the composition of the  showers, the production of scattering tables for each single particle would require an event-by-event simulation and a huge amount of computational time.  
A possible way to solve the problem is to assign  a photon yield equivalent to  that of an electromagnetic shower induced by an electron of a certain energy to particles in hadronic cascades.
\begin{figure}
  \centering
  \includegraphics[width=0.8\textwidth]{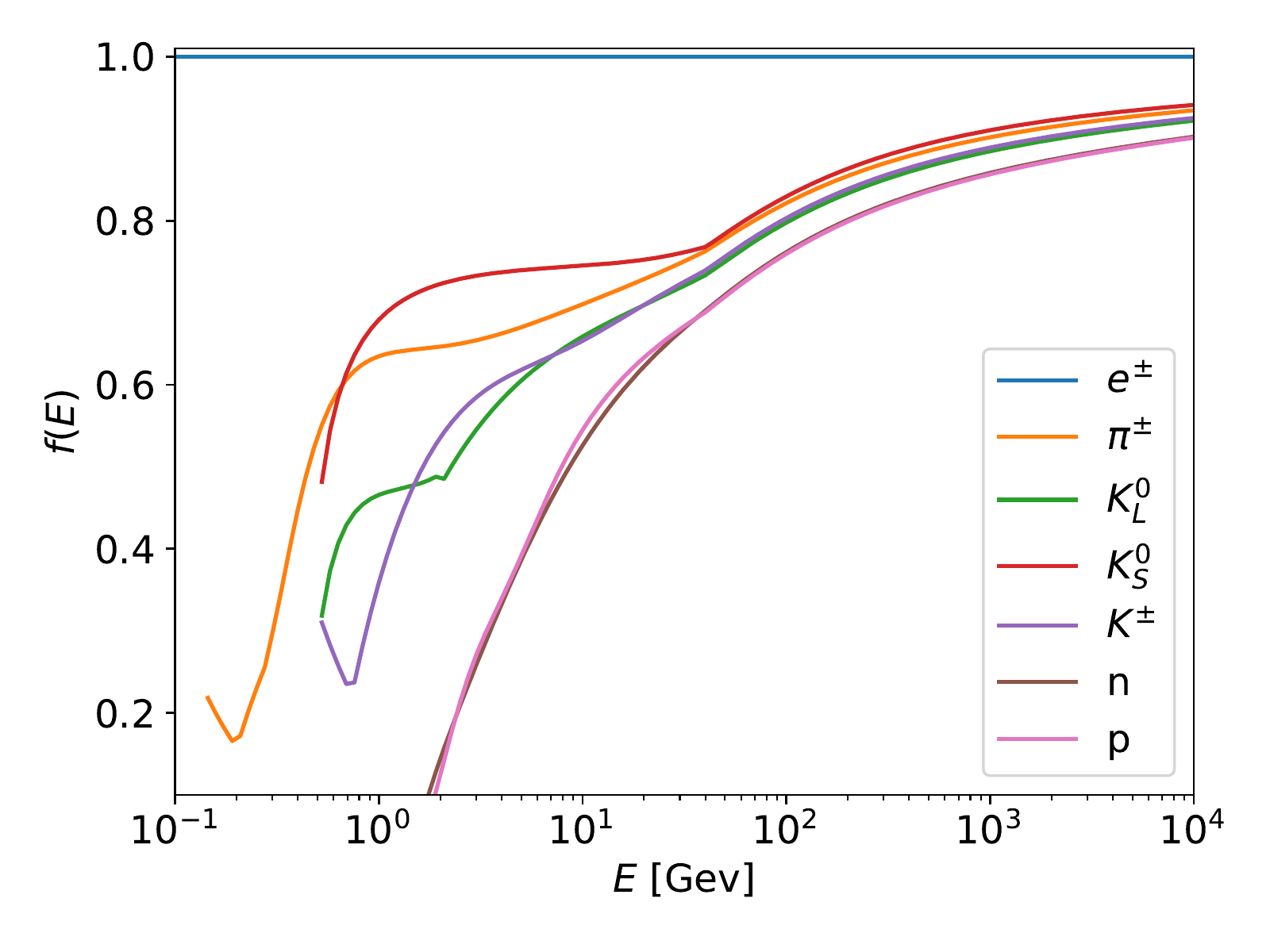}
  \caption{Weight function used in the simulation of particles produced in hadronic showers.}
  \label{fig:mpa}
\end{figure}
A detailed simulation of the light produced by several particles that are present in hadronic showers has been performed.    Assuming  the light production from electrons/positrons equal to 1, the weight {\it{f(E)}} to be assigned to each particle, depending on its energy, is shown in Figure \ref{fig:mpa}. The main limitation  is constituted by the details of the shower shape at low energies and  at short distances ($< $ 4 m), between the vertex of the shower and the PMT. In general, considering the high energy of the events detected by ANTARES and the average  photon track length  to the optical modules that is much longer than 4~m, 
the detailed spatial modeling of the shower is not relevant for the event signature in the detector and  this approach is a good tool to describe hadronic shower light production.
\section{Data acquisition simulation} \label{sec:daq_sim}
The last step of the simulation chain aims at transforming the list of hits on the PMTs into a data stream with the same format and environmental conditions as real data. In order to meet this objective the environmental optical background must be considered and added to the light produced by physics events.
Also single OM behaviour can be affected by  local changes of environmental conditions. As a consequence, the time evolution of the data acquisition must be properly reproduced.
\subsection{ $^{40}$K decay and bioluminescence}
The deep sea is not entirely a dark place and the contribution to the counting rate due to the environment is significantly variable with time. 
The  decays of $^{40}$K dissolved in water represent an almost constant source of light in seawater, which is registered by the OMs.  The main decay processes are:
\begin{eqnarray}
 \nonumber \rm {^{40}K\rightarrow\,{^{40}Ca} + e^{-} + \bar{\nu}{_e}}\\
 \nonumber   \rm { ^{40}{K+e^{-}\rightarrow\,{^{40}Ar + \nu_e + \gamma}}} .
  \label{eq:k40}
\end{eqnarray}
The energy of the electron in the first process can exceed  the Cherenkov threshold in water. The $\gamma$-ray from the second reaction has an energy of $1.46$~MeV and can  induce electrons above the Cherenkov threshold via Compton scattering. 

Biological activity in the deep sea is an additional source of  light even at the ANTARES depth. Its contribution depends on the sea currents and follows a seasonal behaviour, with an observed enhancement in the spring time \cite{bib:current1, bib:current2}. The mean single rate of hits due to the $^{40}$K decays and to biological activity is estimated to be about 50-60\, kHz on a 10-inch PMT \cite{bib:biolum}. In periods of high bioluminescent activity the rate increases, though usually without compromising the detector data taking. In rare and exceptional occasions the measured light rate can reach the MHz level. In this case, the data acquisition is suspended in order to avoid damage to the PMTs. Also bioluminescent bursts can occasionally occur, temporarily overwhelming individual OMs. \subsection{Read-out and trigger simulation} 
During standard data acquisition, all hits  recorded by the PMTs are transformed into  digital signals by the front-end boards  and transmitted to shore without any  selection. This is the so-called  ``all-data-to-shore" approach \cite{bib:antares_daq}. Once on-shore, the data are handled by filter algorithms looking for signals embedded in background. The applied filter algorithms are adjusted according to the detector conditions, e.g. bioluminescence level.
The creation of a reliable simulated data stream requires the extraction of the environmental conditions (optical background) and filter algorithms directly from the real data. In order to accomplish this task, a ``run-by-run"  strategy (see Sec. \ref{sec:rbr}) has been developed that allows  to correctly reproduce the environmental conditions during the data run. The run-by-run strategy accounts for seasonal variations related to the biological activities and for OM inefficiencies due to the ageing of the PMTs and to the biofouling on the OM's surface.
In order to build a close-to-reality data stream, a time window of  2.2$~\mu$s is opened before and after the hits produced by the considered physics process (neutrino interaction, atmospheric muon) and in this time windows the expected number of background hits are added. After that, the effect of electronics is simulated. 
In real acquisition, a front-end Analogue Ring Sampler (ARS) chip integrates \cite{bib:antares_electronics} the analogue signal from the PMT over a time window of 25--30 ns. After the integration, the ARSs have a dead time of  about 250 ns. A second ARS, connected to the same PMT, digitises signals arriving afterwards. 
The so-called \textit{level zero} (L0) trigger selects hits with a greater charge than a predefined threshold, typically set at 0.3 p.e, and send them to shore. The \textit{first level} trigger (L1) is built up, at shore level, of coincidence hits in the same storey within a 20\,ns time window or of a single hit whose charge amplitude is greater  than a tunable ``high threshold'',  between 2.5 p.e. and 10 p.e. A trigger logic algorithm, a \textit{level 2} trigger (L2), is then applied to data and operates on L1 hits. 
Finally, following the same procedure adopted for real data, the physics trigger algorithms used during the acquisition are applied to the simulated data stream and potentially interesting events are stored and processed with the reconstruction programs used for real data.

\section{Run-by-run approach and time evolution of the optical modules efficiency} \label{sec:rbr}

Under the sea, environmental conditions suffer significant variations on different timescales and  directly affect data acquisition in a neutrino telescope like ANTARES. Biological phenomena show evolving trends  producing a seasonal change of the rates registered at the detector. Modifications on a shorter time scale are also present as the change of the sea current velocity modifies the optical rates  \cite{bib:current1, bib:current2}. In addition, not all detector elements take data continuously, because of temporary or permanent malfunctioning of optical modules or lack of connection to some parts of the apparatus, occasionally producing no signal from some ARSs. Finally, environmental conditions affect the choice of the trigger algorithms that are applied during the onshore processing of the raw data stream.
In order to reproduce the detector response under the specific conditions of each individual run, whose typical length is a few hours, neutrino interactions and atmospheric muons are simulated following a strategy described below and denoted as {\it run-by-run}.

First, the temporarily or permanently non-operational OMs are masked in the simulation.
Secondly, the optical background, which might vary due to bioluminescence or bursting activity, is extracted directly from the data considering short segments of the data stream (the {\it frame}, about 104 ms long). Each data frame provides the L0 average rates on each OM, a value that is used for simulating the optical background according to the measured distribution of the hits.
As a result, the effects of  temporary interruptions of data transmission,  of nonfunctioning PMTs, etc. are automatically reproduced in the simulation. 
A connection to the database interface allows to retrieve information on the data acquisition status of each detector element, on the active trigger in each run and on the detector configuration  together with the information about the alignment of the PMTs \cite{bib:positioning}, their individual position and orientation, and  time and charge calibrations.

\begin{figure}
  \centering
  \includegraphics[width=0.6\textwidth]{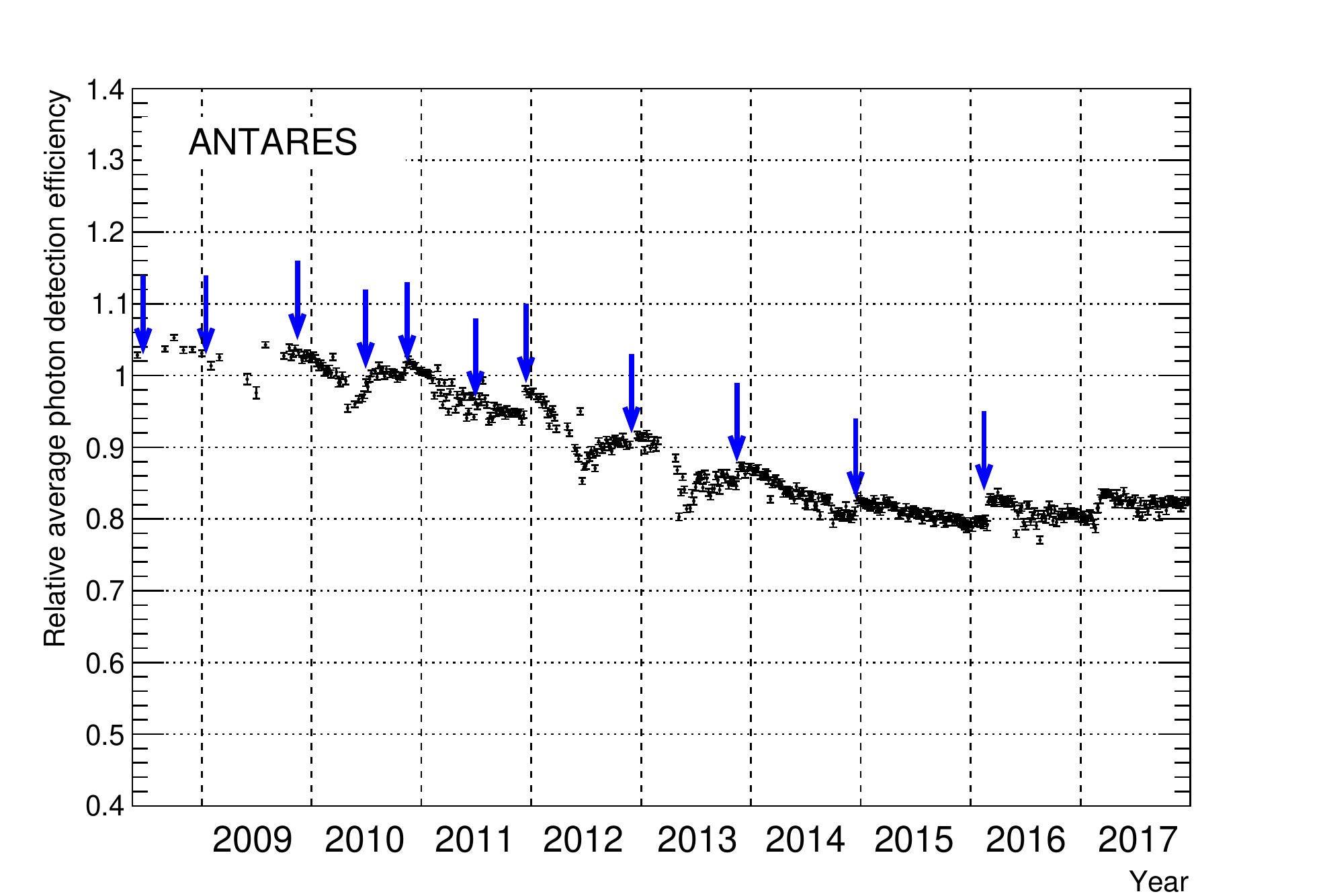}
  \caption{Relative OM efficiency averaged over the whole detector as a function
of time.  The blue arrows indicate the periods in which high voltage tuning of
the  PMTs  has  been  performed,  while  error  bars  indicate  the  statistical  error on the mean efficiency \cite{bib:k40}. }
  \label{fig:k40}
\end{figure}

Thirdly, other inefficiencies on longer time-scales (in particular, on OM efficiency and PMT gain) can be taken into account by feeding the program with specific input files arising from the measurement of the $^{40}$K decay rate.
The signal rate due to the $^{40}$K decay is constant and stable and can be used as a  calibration tool. The Cherenkov light due to the decay of a $^{40}$K nucleus can be registered in coincidence by two OMs on the same storey and the rate of coincidences used as a reference to monitor the OMs efficiency of photon detection. Figure  \ref{fig:k40}, extracted from Ref. \cite{bib:k40}, shows the evolution of the OMs efficiency with time over the period 2008--2017.
The reference value of OM's efficiency, $\epsilon$=1, corresponds to a coincidence rate of 15 Hz, which is the  value obtained with simulations when the  OM properties and the expected decay rate of $^{40}$K are considered. The blue arrow corresponds to the periods when high voltage tuning was performed on the PMTs. With this procedure  the effective gain of individual PMTs is maintained at the level of the nominal gain. 
On average, the detection efficiency of the OMs has decreased by 20\% in the considered period, and tends to saturation.  
The efficiency decrease is not due to the ageing of the PMTs  only. In fact, the correction to inefficiency evaluated  with the method based on the $^{40}$K decay  can reproduce the time evolution of the atmospheric neutrino flux rate and of other features, but is not able to fit the  atmospheric muon rate. An additional correction is required and this suggests that other effects might play a role, for example a larger sedimentation \cite{bib:sed} on the upper part of the OMs that would affect  the detection of light due to the atmospheric muon flux more than in the case of upward going tracks. The efficiency loss in the case of atmospheric muons has been parameterised starting from the data and a correction to the simulation applied. The ratio between the  average rate of atmospheric muons measured with ANTARES and the expected rate simulated with MUPAGE after the application of the OM efficiency corrections is contained within $\sim \pm 10\%$. 

At the end of the full chain of simulation with the run-by-run strategy,  a set of  files is available for each run of the real data acquisition. They are processed with the same reconstruction algorithms and analysis procedures used for the corresponding data files.
The simulation of neutrino interactions is split in  two different energy intervals: low energy regime, 5 GeV to 20 TeV, and high-energy regime, 20 TeV to 100 PeV.  For each energy interval, simulations are performed separately for $\nu_e$, $\bar\nu_e$ and $\nu_\mu$, $\bar\nu_\mu$ interacting through charged and neutral currents.
When $\nu_\tau$ and $\bar\nu_\tau$ are considered, three different cases are treated: NC, CC with a tau decaying into a muon and CC with a tau decaying to an electron or hadrons and thus inducing a  cascade of particles. All these files are combined applying a proper weighting procedure, in order to have either an atmospheric neutrino flux or a neutrino flux with specific features (spectral index of the energy spectrum, energy cut-off, etc.).
Moreover a file for atmospheric muon simulation is produced, whose  livetime corresponds to a fraction  of the real run livetime, usually 1/3.
\section{Conclusions} \label{sec:conclusion} 
The main steps of the MC simulation procedure used in the analyses of data collected with  the ANTARES neutrino telescope are presented and discussed. The peculiarities of the marine environment and the variations of the contribution to the optical background  require special care to follow up and reproduce the time evolution of  the data taking conditions. Thanks to a  procedure that extracts ongoing information directly from the real data (the run-by-run simulation), the MC sample produced so far represents a reliable tool for all ANTARES physics analyses. Though the details of the simulation are strictly connected to the installation site of the detector,  to its properties, and to the geometry of the OMs, the general scheme is valid for any other underwater detector, in particular for the two next generation KM3NeT telescopes: ARCA, for high-energy astrophysics and ORCA for lower energy neutrinos and particle physics \cite{bib:LoI}. 
\section*{Acknowledments}
The authors acknowledge the financial support of the funding agencies:
Centre National de la Recherche Scientifique (CNRS), Commissariat \`a
l'\'ener\-gie atomique et aux \'energies alternatives (CEA),
Commission Europ\'eenne (FEDER fund and Marie Curie Program),
Institut Universitaire de France (IUF), LabEx UnivEarthS (ANR-10-LABX-0023 and ANR-18-IDEX-0001),
R\'egion \^Ile-de-France (DIM-ACAV), R\'egion
Alsace (contrat CPER), R\'egion Provence-Alpes-C\^ote d'Azur,
D\'e\-par\-tement du Var and Ville de La
Seyne-sur-Mer, France;
Bundesministerium f\"ur Bildung und Forschung
(BMBF), Germany; 
Istituto Nazionale di Fisica Nucleare (INFN), Italy;
Nederlandse organisatie voor Wetenschappelijk Onderzoek (NWO), the Netherlands;
Council of the President of the Russian Federation for young
scientists and leading scientific schools supporting grants, Russia;
Executive Unit for Financing Higher Education, Research, Development and Innovation (UEFISCDI), Romania;
Ministerio de Ciencia e Innovaci\'{o}n (MCI) and Agencia Estatal de Investigaci\'{o}n:
Programa Estatal de Generaci\'{o}n de Conocimiento (refs. PGC2018-096663-B-C41, -A-C42, -B-C43, -B-C44) (MCI/FEDER), Severo Ochoa Centre of Excellence and MultiDark Consolider, Junta de Andaluc\'{i}a (ref. SOMM17/6104/UGR and A-FQM-053-UGR18), 
Generalitat Valenciana: Grisol\'{i}a (ref. GRISOLIA/2018/119), Spain; 
Ministry of Higher Education, Scientific Research and Professional Training, Morocco.
We also acknowledge the technical support of Ifremer, AIM and Foselev Marine
for the sea operation and the CC-IN2P3 for the computing facilities.

\end{document}